\documentclass[%
 reprint,
superscriptaddress,
noshowpacs,
nofootinbib,
 amsmath,
 amssymb,
 aps,
 longbibliography,
 pra,
floatfix,
]{revtex4-1}

\usepackage[utf8]{inputenc}
\usepackage[most]{tcolorbox}
\def\Abhilash{Abhilash}
\def\Nicole{Nicole}
\definecolor{light-gray}{gray}{0.95}
\newtcolorbox{Email}[4][]
{
    colframe    = light-gray,
    colback     = white,
    coltitle    = #2!20!black,
    title       = \textbf{From:} #3\newline \textbf{To:} #4,
    #1,
}

\usepackage{scrextend}

\usepackage{graphicx}
\usepackage{dcolumn}
\usepackage{bm}
\usepackage[margins]{trackchanges}

\usepackage [english]{babel}
\usepackage [autostyle, english = american]{csquotes}
\MakeOuterQuote{"}

\begin{document}

\preprint{APS/123-QED}

\title{Operationalizing relevance in physics education: using a systems view to expand our conception of making physics relevant}

\author{Abhilash Nair}
\affiliation{Department of Physics \& Astronomy,Michigan State University, 567 Wilson Rd, East Lansing, MI, 48824}
\author{Vashti Sawtelle}
\affiliation{Lyman Briggs College, Michigan State University,919 E Shaw Ln, East Lansing, MI, 48824}
\affiliation{Department of Physics \& Astronomy,Michigan State University, 567 Wilson Rd, East Lansing, MI, 48824}

\date{Received-\today}
\begin{abstract}
A common hope of many physics educators and researchers is that students leave the course with a stronger sense that physics is relevant to them than when they entered the course. Multiple survey measures have attempted to measure shifts in students' beliefs on the relevance of physics but frequently the results show a negative shift in students' beliefs and are often reported as a failure of students to "see the relevance." We challenge this interpretation by first operationalizing relevance as a construct by using existing theories on beliefs and attitudes. We utilize ecological systems theory to identify rich sites of relevance in students lives and present evidence to demonstrate the rich ways students are able to make relevant connections to physics. We then reflect on the implications of this expanded view on the limitations of past measures of relevance. We articulate how incorporating students' disciplinary ideas and expertise into the classroom can challenge previous deficit-framed narratives of students' abilities to find the relevance of physics. 
\end{abstract}

\maketitle

\section{Introduction}
National policy recommendations for the integration of knowledge across the disciplines continue to position physics concepts and reasoning skills as being important and useful to degrees in STEM or careers in the health sciences \cite{AAMC-HHMI2009,Quinn2011}. Many undergraduate students pursuing a degree in STEM will be required to complete an introductory physics course \cite{Sadler2001}, but research in Physics Education Research (PER) suggests that students do not share the belief that physics is relevant to them \cite{Redish1998a,Brewe2009,Adams2006,Kortemeyer2007,Moll2009,Perkins2006,Cahill2014,Slaughter2012,Halloun1998,Marx2004}. 

In a typical large physics department, non-physics majors constitute one of the largest proportion of students in an introductory physics classroom (algebra-based and calculus-based combined) \cite{AIPEnrollment1997}. It is important to the success of these students, who do not plan on pursuing physics as a career, that they are able to build connections from physics to their lives \cite{Elby2001a,Redish1998a}.

Our purpose in this paper is two-fold --- first to review the ways in which scholars in PER have attempted to probe students' sense of relevance and show how the picture of relevance generated by such measurements is incomplete. Our second goal is to operationalize a construct of relevance that explores the systems that compromise students' lives and use it to analyze student experiences in an introductory physics for the life sciences course. We present the experiences from two case studies that highlight the affordances of this approach in capturing a richer, more expansive picture of students' sense of the relevance of physics. During this work, we critically examine the implications of having an incomplete picture of students' abilities to connect physics to their lives in perpetuating deficit-interpretations of the abilities of life science majors.

\subsection{What we know about relevance}
Even in the midst of calls for instructors to make their curricula more relevant to students, there is much disagreement across the disciplines on the definition of relevance \cite{Bookstein2007,Harter1992,Newton1988}. Newton (1988) noted that "the notion of relevance in science education seems fraught with inconsistency, obscurity, and ambiguity." Bookstein described this issue very concisely \textemdash "relevance is one of the most central concepts of information retrieval; however, attempts to provide it with a definition have been frustrating and confusing" \cite{Bookstein2007}. Scholars of relevance agree that the sense of relevance we are exploring is a relationship between the participant and the subject in question \cite{Bookstein2007,Newton1988}. In order to study this sense of relevance in our physics classroom, we focus on students' experiences that contribute to development of meaningful relationships between classroom elements and their lives.

Relevance is a challenging construct to explore due to the differing meanings in the literature. Relevance has been used as a synonym for student interest, perceived meaningfulness, perceived utility, relating to real-life, or a combination of all the above. Stuckey et al. (2013) trace the differing notions of relevance in science education throughout history and find that there is an evolving focus of policy recommendations on how science \textit{should} impact areas of students' lives. The authors identify three possible purposes for making science education relevant \cite{Stuckey2013}: (1) preparing students for potential careers in science and engineering; (2) helping learners understand scientific phenomena and coping with challenges in life; and (3) supporting students becoming effective future citizens in society.

Although the specific meaning of relevance may be difficult to pinpoint in the literature, we have decades of policy and curriculum recommendations from science educators outlining how relevant instruction should or could impact students' lives. Our goal here is to shed light on how student experiences can inform the ways in which a physics classroom can be relevant. In order to operationalize relevance as a construct, we draw upon the work already done in the PER community around probing and describing relevance. We start by focusing on areas of students' lives that surveys in PER have probed, highlight its limitations, and articulate a more expansive view of relevance that will then serve as a research lens we can apply to students' experiences.

\subsection{How physics education research has probed relevance}
Students' beliefs around the relevance of physics has been a focus of many many attitudinal and epistemological surveys used in physics education research \cite{Adams2006,Redish1998a,elby1998epistemological,Halloun1998}. In this section, we focus on the Colorado Learning Attitudes about Science Survey (CLASS), the Maryland Physics Expectations Survey (MPEX), the Views About Science Survey (VASS), and the Epistemological Beliefs Assessment for Physics Science (EBAPS) to show how they have probed relevance. We argue that the image of relevance generated from the items on these surveys is incomplete.

Starting with these survey measures and expanding to include scholarship around relevant physics instruction, we arrive at areas of students' lives being explored and the desirable outcomes after instruction:

\begin{itemize}
\item Future Career: Students reporting that the physics they have learned will be of use in their planned future career \cite{Redish1998a,Etkina2001,Bennett2016}

\item Real World\footnote{We have argued that this category can be problematic in interpreting students negative responses \cite{Nair2018}}: Students reporting that physics is connected to the world in which they live. \cite{Adams2006,Redish1998a,Halloun1998,elby1998epistemological}

\item Everyday Life: Students reporting thinking of, talking about, or using physics in their daily life \cite{Redish1998a,Adams2006,Halloun1998,elby1998epistemological}

\item{Personal Interest: Students' reporting that they enjoy doing physics or that it provides a sense of satisfaction} \cite{Redish1998a,Adams2006,Halloun1998,elby1998epistemological}
\end{itemize}

Beyond these survey measures, there have been recent efforts to expand what areas of students' lives contribute to their connection with physics, specifically looking at disciplinary interests. Crouch, Geller, and colleagues have investigated student attitudes, interests, and performance in physics activities framed expansively with biological contexts \cite{Crouch2018,Geller2018}. In the next sections we explore each of these surveys in more detail.

\subsubsection{MPEX - Reality Link}
The Reality Link cluster of the MPEX is described as probing whether students believe that ideas learned in physics are relevant and useful in a wide variety of real contexts, rather than having little or no relation to outside experiences \cite{Redish1998a}. The authors of the MPEX directly state that interpretations of this cluster are about the relevance and utility of physics.

In Redish and colleagues' large study using the MPEX, spanning six institutions (N=1,528 students), they report that students entered with strong favorable responses in the Reality Link cluster. By the end, however, "every group showed a deterioration on this measure as a result of instruction, and some of the shifts were substantial \cite{Redish1998a}." This is not uncommon, a decline in favorable responses or an increase in unfavorable responses on the Reality Link cluster has been reported by studies across a variety of curricula, pedagogies, and student compositions \cite{Sahin2009,Crouch2001,Kortemeyer2007,VanAalst2000,Sharma2013,Wutchana2011}. The Reality Link cluster was intended to probe whether students believe that ideas in physics are relevant and useful to a wide variety of real contexts rather than having little or no relation to outside experiences. The implication being that students with a negative shift in this cluster do not believe ideas learned in their physics course are relevant to real contexts or their experiences.

\subsubsection{CLASS - Personal Interest \& Real World Connection}
The CLASS has two clusters that appear to probe the relevance of physics, Real World Connection and Personal Interest. These clusters are described as probing whether students find the physics ideas they learn to be interesting or connected to the real world. \cite{Adams2006}

The authors of the CLASS have reported that the typical result for both the Personal Interest and Real World Connection clusters is a negative shift in favorable responses \cite{Adams2006}. Negative results on these two clusters were reproduced in several other studies as well \cite{Moll2009,Perkins2006,Cahill2014,Slaughter2012}. 

There have been some positive results reported on this cluster; Zhang and colleagues conducted a study of 441 students and compared CLASS results across traditional lecture methods and Peer Instruction. They report traditional methods resulting in negative shifts in both clusters and Peer Instruction sections producing generally positive in both clusters \cite{Zhang2017}. Brewe and colleagues initially reported positive shifts in both clusters in a small study of two semesters (N=22 and N=23) of physics taught using Modeling Instruction and were able to reproduce positive shifts with a larger sample of 221 students \cite{Brewe2009,Brewe2013}. Another study of 44 students utilizing Modeling Instruction by de la Garza and colleagues reported a null shift in the Personal Interest cluster but a positive one in the Real World Connection cluster \cite{DelaGarza2010}. From these studies we see that the typical result for the CLASS across many classroom contexts is negative, with some examples of positive results. 

\subsubsection{VASS - Personal Relevance}
The VASS has had a personal relevance cluster\footnote{Personal Relevance appears in VASS version P204 and is absent in P05.07} consisting of five items that were intended to probe if physics was relevant to everyone's life rather than the exclusive concern of scientists and if studying physics was an enjoyable experience \cite{Halloun1998}.

The designers of the VASS state that "traditional physics instruction has no significant effect on student views about science" and that "on most VASS items, students tend to shift a little more toward folk [considered opposite of expert] views than expert views, after instruction." Unfortunately we could not find studies that reported shifts from pre- to post-instruction within the Personal Relevance cluster.

\subsubsection{EBAPS - Real-life Applicability}
The EBAPS contains a cluster called Real-life Applicability, which probes whether students believe that the ways of thinking in a physics class is restricted to the classroom or if it is applicable in real life. In other words, do students find the ways of thinking in a physics class to be \textit{\textbf{relevant}} outside the classroom?

In studies of high school physics students in California (N=27) and Virginia (N=55) Elby (2001) reports a null shift in this cluster in the California study, and a positive shift in the Virginia study \cite{Elby2001a}. He reflects that the "failure in California caused me to use more real-life examples and to make other modifications when I taught in Virginia." In a larger study (N=255) Marx and colleagues created a modified cluster of items from both the EBAPS and the MPEX probing the relationship between classroom science and the real world. In both the traditional course as well as the "learning-centered" course using research-backed practices, they find a negative shift in this cluster. The authors reflect on this finding, "our Learning-centered course implicitly addresses issues related to several of the clusters by... having students experience simple, explainable, real-world phenomena in the classroom (Reality). Nevertheless, it fails to improve students' attitudes \cite{Marx2004}."

Clusters in these four surveys in physics education claim to probe student beliefs of the relevance of physics. It is the results from these surveys along with other scholarship around how students interpret the role of physics, that have formed physics education's understanding of the question: Do students believe physics is relevant to them? When exploring the results to this question, we find that negative shifts in beliefs around the relevance of physics are typical but that positive shifts are possible. It remains unclear how to use these results to inform changes to physics courses to directly impact students' beliefs around the relevance of physics. In the next section, we will reflect on some of interpretations scholars have made about students' unfavorable beliefs around the relevance of physics.

\subsection{Deficit-based interpretations of student beliefs around relevance}

One of the challenges of probing student beliefs around relevance is that the endorsements of belief statements are often placed on a spectrum between novice-like and expert-like. Unsurprisingly, experts in physics believe that physics \textbf{\textit{is}} relevant. The gap between students' novice-like beliefs and experts' beliefs can sometimes to lead to the problematic interpretation that students with unfavorable responses have not committed themselves to making connections or have failed to see the relevance of physics. For example, Kortemeyer reported in his study of pre-medical students that "the results of the MPEX indicate that over the course of the semester, the perceived relevance of physics actually decreases. \citep[p. 3]{Kortemeyer2007}" Reflecting on this finding, he notes "contrary to the student responses in the MPEX Reality Link Cluster, physics simply is relevant for a physician. \citep[p. 3]{Kortemeyer2007}"

The notion that students are somehow lacking the ability or unwilling to see the relevance of physics is reflected in scholarship that suggest instructors should persuade, demand, or force students to make connections. "In order to realize instructionally significant gains in epistemology, it seems we must carefully craft materials demanding students to overtly and critically evaluate how they learn science and the nature of science itself. \citep[p. 4]{Marx2004}" Bennett, Roberts, and Creagh (2016) reported the success in establishing relevance in a foundational physics course through a 2-hour workshop focused on self-reflection and group work designed to position students' learning in relation to their future lives and careers. They recommend instructors "ensure that students perceive the material they are asked to learn as authentic and of relevance to their future lives and careers. \cite{Bennett2016}" 

We argue that these recommendations should be modified to move the focus away from fixing the student and toward designing classrooms to invite and support students in bringing-in their experiences, thereby incorporating the students into what it means to learn physics. Our view is in line with and supported by a study by Gray and colleagues (2008) that found that students reporting novice-like beliefs on the CLASS are "quite aware of what physicists believe about physics and learning physics; they just do not believe that these ideas are valid, relevant, or useful for themselves. \cite{Gray2008}" They recommend that instructors should should concentrate on strategies "that go well beyond telling students about how experts view physics and focus on making adoption of expertlike views truly useful and relevant for students. \cite{Gray2008}"

Life science students taking a physics course may only be required to take one or two semesters of physics depending on their major. If physics education continues to position student beliefs on a spectrum of expert-like and novice-like, or on a degree of "sophistication", we risk underestimating the relevant connections life science majors are able to make. Adams and colleagues have found that "students' incoming [CLASS Personal Interest cluster] scores increase with level of physics course. Thus, students who make larger commitments to studying physics tend to be those who identify physics as being more relevant to their own lives \cite{Adams2006}" We hope to complicate this picture by providing examples of life science majors without a "larger commitment" to studying physics beyond the introductory sequence who articulate relevant connections to physics. 

We argue that insisting that physics, as it is taught, \textbf{\textit{is}} relevant is bound to create tension with students due to mixed messaging they may receive on the relevance of physics. Beverly (pseudonym), who is majoring in human biology with the intention of attending medical school, states that physicians she has shadowed have told her they do not find physics relevant to what they do.
\begin{addmargin}[.45cm]{.45cm}
\textit{I talked to like various physicians I've shadowed and asked them like what do you actually use... they're like... "I don't really use that much physics"} [Beverly, Interview 1]
\end{addmargin}
Beverly's belief that physics is a course she just has to get through seems in agreement with the beliefs of medical students, graduates, and physicians who state that physics was irrelevant to their success \cite{Dienstag2008,Eyal2006,Pabst1997}. There have also been studies reporting that physics has no strong correlation with success in medicine \cite{Muller2010,Hall2014}.  We will show how, instead of challenging students' beliefs about the relevance of physics to their future careers, we can start to incorporate their rich disciplinary experiences to make physics more relevant.

Failure in a student's ability to see physics connecting to one's world or life appears to be damning indictment of the abilities of students to engage in meta-cognition about their coursework. We don't believe the authors of every article reporting negative results on clusters from these surveys mean to suggest this. We believe the issue lies in the limited picture of relevance and what it means to \textit{measure} a student's sense of relevance. We suggest that perhaps the measure of relevance provided by such attitudinal and belief surveys is incomplete, especially with regard to how disciplinary experiences outside of physics impact that sense of relevance. We present analysis in this paper that pushes against deficit interpretations and instead argues that if students' ideas are brought into the classroom, students can and do make relevant connections.  

\subsection{An incomplete picture of relevance}
We argue that results from clusters probing relevance are not sufficient to form a complete answer to the question: Do students believe physics is relevant to them? The majority of the work on relevance in PER \cite{Adams2006,Halloun1998,elby1998epistemological,Redish1998a,Bennett2016} has been centered on the local context of the physics classroom, how physics may impact students' future careers, or how students relate physics to their world (the generalized "real world" or their everyday lives). These areas of intersection between the physics classroom and a student's life fail to capture some critical ways in which students inform their attitudes toward physics. Students often arrive at the introductory physics classroom with attitudes and beliefs informed by experiences outside of the physics classroom, conversations with friends and family, and the disciplinary perspectives they may hold. We argue for expanding our notion of what contributes to a student's perception of the relevance of physics to include these additional experiences. The students in this article are all life science majors; they have a rich set of disciplinary experiences outside of physics that may inform and impact their relationship with physics \cite{Sawtelle2016}. Through their experiences, we will show that expanding our probes to include disciplinary experiences outside of physics captures a richer image of how physics can be relevant.

To support this expansion of what contributes to relevance, we adapt ecological systems theory \citep{Bronfenbrenner1979} to represent students as existing in overlapping systems that all contribute to their perception of the relevance of physics. This approach enables us to build a construct of relevance that goes beyond treating a student's sense of relevance as contained \textbf{\textit{within}} the student, it allows us to ask questions about the intersection of the many experiences in a student's life that have contributed to their  view of the relevance of physics.

\section{Theoretical Framework}
In this section, we revisit the messiness of relevance as a construct and attempt to give it more structure and definition. In constructing a theory of relevance for physics education research, we remind ourselves of the purpose of this entire enterprise. We are interested in finding out if we, as instructors and researchers, can impact students' beliefs around physics. With this in mind, we set out to answer the following questions. For what purposes will students retrieve the information they have learned in physics outside of the classroom or vice versa? Will students find meaning or value in physics beyond its purpose in the classroom? Do students construct productive connections between physics and other scientific disciplines?

\subsection{Positioning relevance as a theoretical construct amongst attitudes \& beliefs}
"Do you \textit{\textbf{believe}} physics is relevant to you?" How we colloquially probe relevance in conversations in invokes belief systems, as does the language used in survey measures in PER \cite{Adams2006,elby1998epistemological,Redish1998a,Halloun1998}. Beliefs, then, serve as a starting theoretical foundation for us to build upon in operationalizing relevance. Pajares provides a thorough view of beliefs, tracing its development as a unit of analysis and the challenges in arriving at a consensus for the meaning of a belief. He articulates a set of findings, inferences, and generalizations researchers have confidently reported about beliefs. Drawing from Pajares's 16 fundamental assumptions about beliefs ---each supported with a body of literature--- and from the decades of focus on improving physics instruction we can move forward reasonably confident that (1) Students' beliefs about physics exist and may impact many aspects of students' lives and (2) students' beliefs about physics are the focus of many efforts to improve physics instruction.

In order to start building a structure to organize relevance, we adapt Rokeach's organization of beliefs and attitudes \cite{rokeach1968beliefs}. Rokeach describes an attitude as a set of beliefs focused on or aligned towards an object or situation \cite{rokeach1968beliefs}. We argue that a student's sense of the relevance of physics is not informed by a single belief, but that a student uses many beliefs, formed and shaped by many experiences, to evaluate how relevant physics is to them. Similarly, we contend that a student can have multiple attitudes that form a sense of relevance. A student may have an attitude around "physics being relevant to their future career" as well as an attitude focused on "physics being irrelevant to their everyday life" and these in concert, along with numerous other attitudes, can influence a student's sense of the relevance of physics.

\begin{figure}
    \includegraphics[width=\columnwidth]{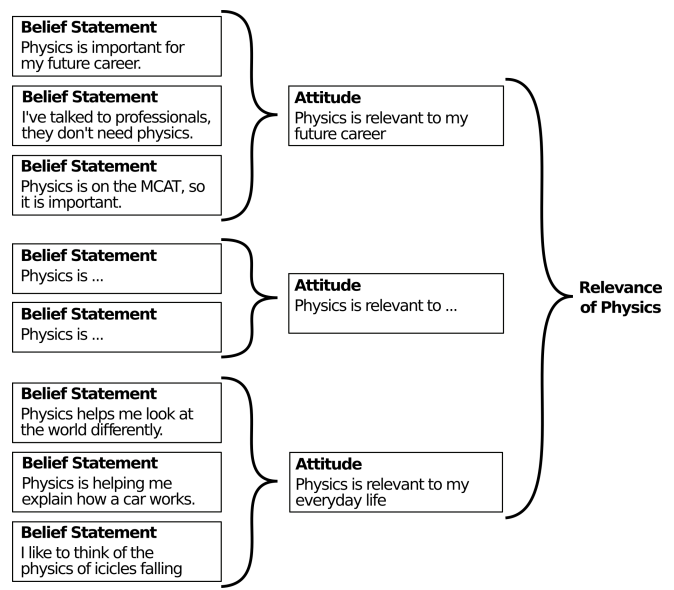}
    \caption{\label{belief-structure}A representation of relevance consisting of multiple attitudes and beliefs. This is not intended to be an exhaustive list of all beliefs and attitudes that help form a sense of relevance, but serves to give an example of such an organization. Multiple beliefs (inferred by belief statements) are organized into attitudes. Multiple attitudes constitute a sense of relevance. Belief statements are used to infer beliefs in this structure as they are the most readily accessible to researchers measuring students' beliefs. This organization of relevance aligns with (1) the structure of previous survey measures of relevance in PER and (2) Rokeach's theoretical organization of beliefs.}
\end{figure}

This organizational structure (Fig \ref{belief-structure}) \textemdash relevance consisting of one or more attitudes, which themselves consist of one or more beliefs \textemdash aligns well with commonly used survey measures in PER \cite{Redish1998a,Adams2006,Halloun1998,elby1998epistemological}. In these surveys, students are asked to endorse individual belief statements by choosing levels of agreement on a five-point Likert scale or to situate their viewpoint between two opposing belief statements on a five-point scale. These individual items representing belief statements are then grouped using quantitative methods to arrive at sets of items or clusters that probe similar ideas. This approach to the design of the surveys aligns with Rokeach's organizational structure: individual question items ask students to endorse belief statements, and a set of these belief statements are grouped together to characterize a broader attitude toward physics.

\subsection{Adapting ecological systems theory}
A student's life has many layers, they participate in a multitude of settings and we expect their sense of the relevance of physics will be impacted by these numerous contexts and interactions. Ecological systems theory (1) preserves the richness and complexity of students' lives, (2) serves as a map for scholars studying relevance to locate relevant intersections of contexts, and (3) provide utility to physics instructors as they attempt to make instruction more relevant to their students. Ecological systems theory was originally developed to characterize different layers of systems that affect human development. Bronfenbrenner (1979) describes the ecology of human development as the study of how a person and the dynamic settings they experience mutually accommodate and adapt to each other under the influence of relations between settings and larger contexts the settings are embedded within \cite{Bronfenbrenner1979}.

There are three underlying features of this framework that will be important in our construct of relevance: (1) The person is considered a dynamic entity with agency to impact their environment; (2) The interaction between a person and their settings is reciprocal by nature, each having the ability to impact the other; (3) The environment that is influencing the person is not limited to a single setting, but expanded to include connections between settings and external influences. We adapt the first two features to mean that a physics classroom does not simply impart knowledge on its students, but rather the classroom is shaped by the students and the experiences they bring \textit{into} the classroom. We adapt the third feature to state that a student's sense of relevance of physics is not solely influenced by their physics classroom.

In adapting these features of ecological systems theory, we bring along a situative view of the classroom\cite{Leach2003,Rogoff1995,Greeno1996}. Specifically, we will consider how the structures and cultural values of a physics classroom can inform relevance through interactions with other settings in a student's life. This situative perspective on what contributes to a student's sense of relevance will have implications for our view of the ability of commonly used attitudinal and belief surveys to probe the whole of relevance. Particularly, items probing students' beliefs around connections between different disciplinary courses are missing. We will show how these connections can be an important contribution to a student's sense of the relevance of physics.

An additional affordance of this situative perspective on relevance is that it gives us the power to describe the enactment of physics classroom with more richness. This view positions the classroom as an actor on the stage rather a passive environment the student experiences. The situative perspective necessitates that we acknowledge that the design and implementation of a classroom is imbued with cultural values through its structures and not simply a collection of curricula and pedagogy. The classroom's tools and materials, discursive practices, participatory structures, and task structures \cite{Sandoval2014} all convey cultural values. We believe that viewing the classroom as one actor among many in a students' ecosystem can help in articulating the structures within the classroom and relationships across classrooms that promote relevance.

\subsubsection{Structure \& Organization of Ecological Systems}
\begin{figure}
    \includegraphics[width=\columnwidth]{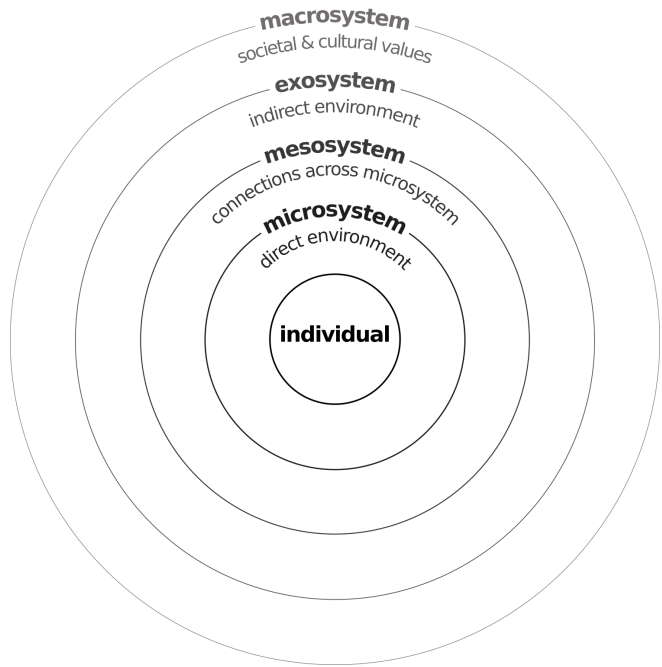}
    \caption{\label{figure-ecological-systems}A visual representation of Bronfenbrenner's \cite{Bronfenbrenner1979} organization of ecological systems theory as a set of concentric systems encompassing an individual \textemdash moving outward from the center are the \textemdash micro-, meso-, exo-, and macrosystems.}
\end{figure}

Bronfenbrenner \cite{Bronfenbrenner1979} organized ecological systems theory as a set of concentric systems encompassing an individual as shown in Fig \ref{figure-ecological-systems}. Each system represents layers of contexts and interactions which may inform the development of the individual. In the descriptions that follow, we will adapt ecological systems theory from its original purpose, which was to explore human development, to describe students in a physics classroom.

The characteristics and identities of the individual including but not limited to age, sex, race, or health are positioned in the center. The first layer encompassing the individual is the microsystem, which is defined as "a pattern of activities, roles, and interpersonal relations experienced by the developing person in a given setting... \cite{Bronfenbrenner1979}"  We can imagine this system being occupied with settings a student actively participates in including their coursework, research groups, jobs, social settings with peers and family, and more. In addition to the settings, the microsystem also includes the roles the students play in these settings and the interpersonal relations within these settings.

The mesosystem "comprises the interrelations among two or more settings in which the developing person actively participates... \cite{Bronfenbrenner1979}" The mesosytem can help us describe the interrelations between settings in the microsystem. For example, a physics course and a biology course may interact with each other through the culture enacted in each course which can incorporate the pedagogy, curricula, or participation structures. A common interaction we encounter in the mesosystem is that our students enter our course having heard horror stories of physics experiences from their peers, family, or from the broader messaging that physics is difficult and not directly applicable to the future careers of life science students. In this paper we will focus on elements of students' mesosystems and show how they can be powerful mediators of a student's sense of relevance.

The exosystem includes the settings that the student does not actively participate in but nonetheless impact or are impacted by the settings in which the student actively participates. Examples of elements in the exosystem include the physics experiences of family or friends, the employment of a student's primary caregiver, or the policies that impact a student's education. This level of the student's ecosystem can also hold future career qualifications, medical school admissions criteria, and even mass media. All of these interactions informs not only a student's participation in the physics classroom, but can also influence the design and implementation of the classroom itself.

The macrosystem contains cultural and societal norms that exist consistently within any set of lower-order systems \textemdash micro-, meso-, or exo- \textemdash and can include belief systems and ideologies that govern the lower-order systems. We can imagine the contributions of the culture and society to a student's perception of the relevance of physics. If we narrow our focus to the culture within specific scientific disciplines, there may emerge a set of norms, attitudes, ideologies, or expectations that govern participation in a variety of settings of a student's microsystem. The macrosystem has important implications for our study of relevance. All of the students described here have a disciplinary major outside of physics; the culture of a student's home discipline can inform their participation in and attitudes toward physics. \cite{Watkins2012,Redish2013}

Bronfenbrenner would later add a layer beyond the macrosystem called the chronosystem which includes events and transitions in one's own life as well as the environment around them. In this study we exclude the chronosystem from consideration as we do not have sufficient longitudinal data from student's entire lives to comment on the significance of major life events and transitions beyond their time in our physics classroom. 

Each of these systems will play a role in students' lives, but for the purposes of this paper we will focus our attention on the microsystem and mesosystem. We believe these two layers represent rich spaces to explore the relevance of physics and are the most readily accessible in the design of this study.

\subsection{Relevance through transformation of participation}
A consequence of using a situative approach to relevance is that the settings and the structures that comprise them are embodied with values and they are intertwined with the co-construction of relevance with the student. Rogoff and colleagues (1995) described this in their work on defining development as a transformation of participation. "Individuals' efforts and sociocultural institutions and practices are constituted by and constitute each other and thus cannot be defined independently of each other or studied in isolation \cite{Rogoff1995}." This directly informs how we interpret students' experiences presented in this paper.  Our students' statements around finding physics to be relevant are often intertwined participatory acts that reciprocally impact and are impacted by the classroom structures. The classroom structures and the student, in concert, shift their participation from experiencing an activity in physics to transforming the activity with their contributions. 

\section{Methodology}
\subsection{Studio \& IPLS classrooms are rich contexts to study relevance}
The context for this work is an introductory physics for the life sciences classroom in a residential college within Michigan State University called Briggs Life Science Studio Physics (BLiSS Physics). This physics course has been recently reformed to leverage connections to biology in the learning of physics. As previously mentioned in the description of the mesosystem, one way we expect students to find relevance in physics is through cross-disciplinary connections. Classrooms that promote building of connections across disciplines are a rich context to explore a construct such as relevance. A student who is majoring in the life sciences may already have a strong sense of relevance of their own areas of study that may not extend into the physics classroom. A context that emphasizes connections between physics and the students' home disciplines provides the opportunity to observe students experiencing and reflecting on these connections. These moments are characterized by the nature of the classroom activity and a student's participation impacting one another.

BLiSS Physics adapts curriculum, discursive structures, and participation frameworks from Modeling Instruction for introductory university physics \cite{Brewe2010,Brewe2008}. In addition to breaking the traditional distinction of lecture and laboratory, our implementation of the studio format in this course emphasizes student-led investigations as an entry point into every unit. Student groups design and implement investigations to arrive at empirical rules supported by evidence that govern a phenomena. The entire classroom then utilizes white-board meetings and argumentation to arrive at a classroom consensus \cite{Brewe2008,Brewe2018}. Computational activities are used throughout the course to support students gaining competency in using vPython to model and visualize phenomena that cannot be done using traditional closed form analytic methods. In these activities, students are presented with minimally working code that runs without errors but is lacking the correct physics. Students then are asked to use their physics concepts to write or correct the few lines of code so that the physics used in the simulation is correct. Students are never asked to write python code from scratch, rather the focus is on students developing an understanding of foundational programming structures (constants, loops, conditional logic statements, etc.) and to gain competency in modifying existing code to suit their needs.

In introductory physics classrooms we often position students as being novice without consideration to the extensive rich disciplinary content knowledge they possess outside of the discipline of physics. To address this, BLiSS Physics intentionally positions students as being expert in their home disciplines and has space designed to allow students to \textbf{bring in} their expertise \cite{Sawtelle2016}. When students engage in bringing in disciplinary content knowledge or life experience into the physics classroom, it presents a significant opportunity to forge connections between students' lives and physics.

\subsection{Data Collection}
The data presented in this article are part of a larger design based research \cite{Bell2010} endeavor in (1) developing and iterating on an introductory physics for the life sciences course and (2) articulating design conjectures \cite{Sandoval2014} in establishing a classroom environment that attends to student affect, positions students as disciplinary experts in the life sciences, and promotes the relevance of physics. Design based research is a situated approach to studying classrooms; the embedded nature of this methodology is reflected in the analysis which coordinates across multiple streams of data across two years of iteration in the course.

Potential case students were identified from a course survey given in the first week of class that asks students to reflect on their previous experiences with physics as well as their disciplinary interests. The survey included both open and closed responses and were reviewed for the presence of certain factors such as students expressing a strong disciplinary identity, high anxiety or fear leading into the first semester of physics, or students who were open to physics potentially playing a role in their future career.

Independent of the survey, students were asked to consent to in-class video recording as well as volunteer to participate in research interviews. Students whose survey responses were deemed interesting based on the factors described above were cross-matched with interview volunteers. A semi-structured interview protocol was developed to probe students' beliefs around the relevance of physics as well as their beliefs around the design of the course. The case students that were interviewed were recorded in class as they worked on activities in small groups. Based on the initial interview, in-class video, and availability students were interviewed multiple times.

Author Nair was embedded in the course both as a researcher and as a member of the instructional team (in sections he was not conducting interviews). Field notes and observations were used to capture significant moments in the classroom as well as the general effectiveness of classroom activities in achieving their learning outcomes. This deep involvement in the classroom environment helped form an understanding of the culture of the classroom as well as the discursive and participatory norms practiced throughout the course.

\begin{figure}
    \includegraphics[width=\columnwidth]{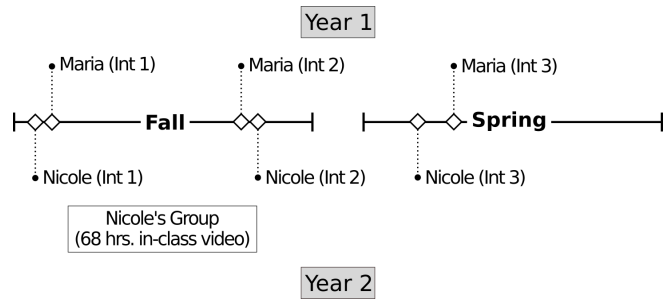}
    \caption{\label{data-colletion}The time-line of data collection for the case studies presented in this paper, Nicole and Maria. Both students took the BLiSS Physics course in the Fall semester and a different second semester physics course in the Spring. In Year 1 of this study, in-class video was recorded of focal groups working through a unit on diffusion. In Year 2, focal groups were recorded throughout the entire semester. This figure represents a subset of a much larger research study that followed over 25 students across the 2 years. }
\end{figure}

The first interview took place in the first few weeks of the Fall semester. The second interview took place approximately halfway through the semester after a unit on diffusion. The third interview was conducted midway through the Spring semester (Fig \ref{data-colletion}). One of the case students, Maria, became an undergraduate learning assistant in the course and was interviewed a fourth time midway through their first semester on the instructional team.

\subsection{Analysis Methods}
In this paper, we include data from two case studies. Maria (pseudonym) was a student in the first year of the course, and Nicole (pseudonym) was a student in the second year of the course. After articulating categories of relevance PER has previously explored (Section B of the Introduction), we conducted semi-structured interviews to probe each category and discover new categories. Each interview with these case students was video recorded and transcribed. We used MAXQDA to code each interview's transcript for evidence of student's beliefs around the relevance of physics. We utilized an open coding scheme to expand and refine areas that contributed to the relevance of physics. Each case student's in-class video was analyzed to find and corroborate events described in interviews, the resulting moments were transcribed and analyzed. In the classroom, author Nair was present for the vast majority of class sessions and recorded field notes of interesting moments in which the case students' senses of relevance may have been impacted. Based on his observations in class author Nair conducted impromptu audio-recorded interviews with the instructor on multiple days to better understand instructional choices and to capture reflections of moments shortly after they occurred. MAXQDA was used to triangulate these different streams of data and to record analytic memos as the project progressed. This analysis was presented and discussed at multiple research meetings to check the validity of interpretations and claims.

In the next few sections, we outline case studies of Maria and Nicole: two students majoring in the life sciences who report relevant connections to physics. Their rich experiences go beyond what can be revealed by current survey measures and challenge the deficit interpretations of life science students' ability to see the relevance of physics. Informed by their experiences, we argue for an expanded view of relevance that includes the connections students make across course structures as well as relevance co-constructed in a classroom designed to support relevant connections. Maria's case highlights the importance of the mesosystem in fostering relevance through connections between courses. Maria's experiences have implications for how a physics course can be designed to foster relevance, but in past scholarship these attempts involve layering-on activities of reflection and meta-cognitive development. Instead of looking for ways we can impose relevance on students, we trace the experiences of Nicole as an example of how course design and participatory structures can impact a student's sense of relevance and have lasting effects after the course.

\section{"Maria"}
Maria is a microbiology major with a minor in epidemiology who identifies strongly as a microbiologist. She has founded or held leadership positions in multiple biology and infectious disease related organizations. She works in a water microbiology research lab, tutors students, and conducts campus tours. At the time of the interview Maria was concurrently enrolled in the physics course and a prokaryotic physiology course. Maria had previously taken physics in high school and recalls that experience as being disconnected from her interests. During this study Maria was in her junior year of her undergraduate education and planned to pursue a graduate degree in public health after graduation. In our first interview, Maria reflects on her previous experience with physics.

\begin{addmargin}[.45cm]{.45cm}
\textit{"The last time I've had physics was sophomore year of high school... I just don't think they did a very good job of connecting it back to everyone's interests...it was just theoretical pure physics. so, not my thing..."} [Interview 1]\\
\end{addmargin}

Maria states that the type of physics course she experienced in high school was not effective in connecting to students' interests in general and has concluded that physics is "not her thing." When asked if she believes physics is relevant to her, she is optimistic about the BLiSS Physics course and suggests that connecting to biology may be a path towards relevance for her.

\begin{addmargin}[.45cm]{.45cm}
\textit{"I think in this course it may be a little more 'cause of the [biology] connections, otherwise I would probably say no [laughs]"} [Interview 1]\\
\end{addmargin}

In this section we explore the ways in which Maria is finding relevance in BLiSS Physics that go beyond what attitudinal and belief survey measures have the ability to capture. We adapt Bronfenbrenner's ecological systems theory to describe some of the contexts in Maria's life that have connected to her physics course and impacted her view of physics. Finally we argue that students like Maria, who are life science majors can and do find physics to be relevant. We shift away from deficit-framing to look for ways in which the design of the physics classroom and the ecology of her disciplinary experiences have supported amplification of her sense of relevance.

\subsection{Describing Maria with ecological systems theory}
If we were to imagine the ecosystem that Maria's life exists within, there may be a multitude of elements that are not visible to instructors or researchers. Even though we are unable to fully describe Maria as a person \textemdash we doubt any theoretical model can fully encapsulate a person's life \textemdash there are many elements and relationships that we are able to see playing an important role in her sense of relevance. We focus on those areas in what follows.

\paragraph*{\textbf{Maria's microsystem:}}
Starting with the most central system, Maria's microsystem includes her courses, her microbiology major program, her water microbiology research lab, her clubs and organizations, as well as her family and peers. Additionally, her microsystem includes the roles she takes on as as well as the interpersonal experiences within these settings. These are the settings in which Maria actively participates and engages in activities and we expect them to contribute to her sense of relevance.

\paragraph*{\textbf{Maria's mesosystem:}}
Maria's mesosystem involves the relations between the settings in her microsystem. Maria was enrolled in a prokaryotic physiology course \textemdash concurrently with the first semester of physics \textemdash which impacted her sense of the relevance of physics in answering questions in microbiology. The research she conducts in water microbiology impacted how she perceives or participates in a physics activity exploring the resistive forces experienced by a water-dwelling paramecium.

\paragraph*{\textbf{Maria's exosystem:}}
Expanding further, the settings in Maria's exosystem include her microbiology major's undergraduate program committee, future graduate schools, and future career opportunities within microbiology. It also includes settings that her family, friends, and peers engage in. These are settings in which Maria doesn't actively participate in or impact, but they have implications for the settings within her microsystem. For example, the requirements in a future graduate program or career can impact the courses that populate her microsystem. Maria's friends and family may have had experiences in classrooms or workplaces that can impact how she participates in her own coursework.

\paragraph*{\textbf{Maria's macrosystem:}}
Lastly, Maria's macrosystem accounts for the culture, beliefs, or ideologies permeating through her more inner systems \textemdash micro-, meso-, and exosystems \textemdash that form a consistent thread throughout her experiences. Maria readily points out that she identifies as a \textit{micro}biologist and not a \textit{macro}biologist. Microbiology as a disciplinary area of study contains a set of norms and expectations on the appropriate scales of research questions as well as ways of communicating within their community. The physics classroom will attempt to set norms and expectations for what it means to do physics, but these norms and expectations interact with and are woven into a larger network of culture, beliefs, or ideologies of what it means to \textbf{do} physics as opposed to biology.

\begin{figure}
    \includegraphics[width=\columnwidth]{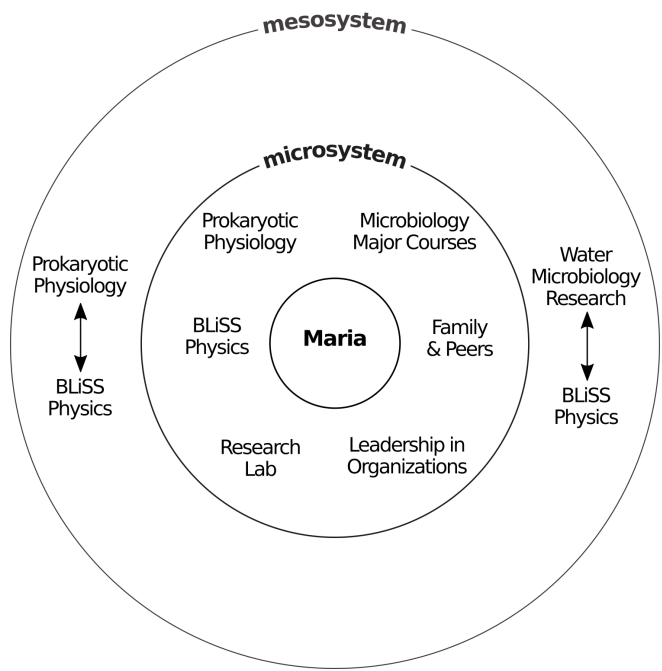}
    \caption{\label{maria}Maria's microsystem and mesosystem. While the whole of Maria's experiences can never be captured in any diagram, this figure represents areas of Maria's life that we consider important to exploring the relevance of physics to Maria.}
\end{figure}

Mapping Maria's experiences onto ecological systems theory helps us see the value of the mesosystem. There are a few items on commonly used PER surveys that probe the mesosystem to ask about potential connections between the physics classroom and the student's life outside of the classroom. We contend that moving away from an abstract sense of \textit{life outside of the physics classroom} and moving toward identifying specific settings will help reveal useful connections instructors and researchers can leverage in building relevance with students. In the next section we dive deeper into Maria's mesosystem for specific settings, roles, and interpersonal relationships that have impacted Maria's sense of the relevance of physics (Fig \ref{maria}).

\subsection{Interactions between physics and microbiology in Maria's mesosystem}
In the first interview with Maria, she was a few weeks into the first semester of physics and already identified potential connections between physics and her other courses. When asked if she saw biology, chemistry, and physics as being related she replied:\\

\begin{addmargin}[.45cm]{.45cm}
\textit{"Why are things moving that way within the cell? Why can the cell do that? Well, it's physics. actually, we talked about lateral motility within the cytoplasmic membrane in um... in microbio, and I'm pretty sure my professor's words were something like 'the thermodynamics and the physics of the you know the cytoplasmic membrane are a glorious chapter in microbio!' {laughs} ... anytime anything is moving, or even just you know the thermodynamic things, things you don't think about, it's physics."} [Interview 1]\\
\end{addmargin}

When asked if she sees connections between physics concepts she is currently learning and her other courses, she notes that the connections are mostly related to motion and that they are basic in nature.\\

\begin{addmargin}[.45cm]{.45cm}
\textit{These are a little more basic but I mean I know it's going to apply somewhere. You know you can find velocity of a bacteria like we did actually in our activity [motion tracking video of bacteria] and stuff like that, and you know how different things affect their velocity and stuff like that. But so far it's pretty basic...} [Interview 1]\\
\end{addmargin}

There are two features of Maria's statements that reflect commonly espoused beliefs in our interviews. (1) Students attribute that many things in the world happen because of physics, or that physics underlies everything. This is in line with the disciplinary culture of physics in which it claims to be fundamental or foundational to other disciplines. \cite{traweek2009beamtimes}  (2) The connections students are most readily able to make early in the course are around motion. Since motion constitutes the first few units of the course \cite{Brewe2008}, it is understandable that students attribute motion of objects or creatures to physics. Maria's early connections do not reveal deeper connections occurring beyond observing that physics sometimes pops up in her other courses in the form of motion or a statement of reverence for the physics underlying more complex mechanisms.

\paragraph*{\textbf{Maria's evolving sense of relevance:}}
Midway through the first semester of physics we interviewed Maria again and ask if she thinks biology and chemistry play a role in her physics activities.\\

\begin{addmargin}[.45cm]{.45cm}
MARIA: Yeah, I laughed. Because so I'm in prokaryotic physiology... I'm a microbio major, the number of times physics comes up is a little... Yeah. Because we were talking about the movement of flagella. And how some parts are using the torque and what does the work, and different things like that. I'm like, this is physics. And then we were talking about the strength of the cell wall and the different components, and what's resisting and the inside forces, and what's really providing the protection in terms of physics. So we never go into details, but the words are like, it's there. You know it's physics.\\
I: And then you're seeing that connected here?\\
M: Yeah, I think when we talked about the diffusion lab, that we had talked about that previously. In that class, the diffusion across the cell membrane, and you know, viewing it as smaller elastic collisions. [Interview 2]\\
\end{addmargin}

Maria is seeing physics come up in her prokaryotic physiology class enough to warrant her laughing about it being a common occurrence. In this exchange we see Maria identifying physics concepts that play a role in her prokaryotic physiology including topics of movement of flagella, forces inside the cell, as well as diffusion across a cell membrane. Later in the same interview, we ask Maria her thoughts on an activity calculating the resistive forces that a paramecium experiences in water \cite{Redish2014}. This activity involved making assumptions about the density of the paramecium and Maria shares a story of her bringing in her microbiology expertise.\\

\begin{addmargin}[.45cm]{.45cm}
Interviewer: How about the paramecium one?\\
Maria: I like the paramecium one.\\
I: You liked it.\\
M: Because I'm a microbiologist. So I was like, this makes so much sense. Because I think actually she [the professor], when you had to take the density of it, I thought of it in a way like, she actually had me tell it to the class. Like she hadn't thought of it that way. But I know as a microbiologist, the paramecium is going to take in its surroundings, like osmosis. And a cell is 99\% water as it is. So I knew I could take the paramecium's density approximately as water. Which is not how, she saw it as like oh, it's floating in the water, so the densities have to be relative. But I would never have thought of it that way. [Interview 2] \\
\end{addmargin}

Maria reflects on her identity as a microbiologist and the disciplinary expertise that she brings with her into the physics classroom. She is able to offer up an alternative reasoning path that differs from the professor's explanation. In the course, positioning students as experts in biology is a critical design feature and it manifests here as the professor validating her idea and having her share it with the entire class. This participatory act of Maria bringing in and successfully incorporating her biology content knowledge into a physics activity is an observable display of relevance.\\

These interactions in Maria's mesosystem between her biology and physics courses prove to be a powerful influence that Maria articulates near the end of the second interview. Maria is asked to expand on why she believes physics has the power to explain why things happen.\\

\begin{addmargin}[.45cm]{.45cm}
Maria: I've always kind of thought of physics as more of a conceptual, being able to explain things. Because it takes it to such a simple and mathematical, you can model with it. But not really until taking this class and getting the tools. Sure, you can see something and be like oh, yeah, physics probably explains that. But I don't know physics, so why would I think about it that way. I don't have that tool. But I think taking this course, the microbiology course alongside physics, where things like work and torque and force are coming up, in a field that I know about. That helps you see... But, I think if I were to look at physics, I think I'm interested in how it relates to the macroscopic biological world. I know other people do other things, but this [paramecium activity] is the physics I like. [Interview 2]\\
\end{addmargin}

In the exchange above Maria states that it wasn't until she was taking the physics class alongside prokaryotic physiology course that she saw how the the tools of physics connect with a discipline she has expertise in. Maria sees connections across the two courses, and states that the physics activity on paramecium represents the physics she would like to do. We believe this is a significant marker of relevance. Maria expresses interest in physics and she sees physics as a tool that can be used in microbiology.

When this physics course was designed, we did not imagine or plan for this synergistic relationship with prokaryotic physiology. It is not feasible for an instructor to predict that a student like Maria will enter their classroom and make significant connections to a specific disciplinary course. We do not credit the relevant connections Maria is able to build to the physics curricular materials or activities, we instead point to designed course structures that make space for students to bring themselves into the physics classroom. Maria brings into the physics classroom her disciplinary expertise and her identity as a microbiologist. The connections she makes are validated by the instructor and amplified by sharing them with her classmates. The physics course structures and environments are not changing Maria, she arrived in the physics course seeking to make connections and to incorporate knowledge across the disciplines. Early in the course, Maria states that she sees physics playing a role in a variety of biological contexts. Through interactions between her courses, we see an amplification of Maria's sense of relevance. We articulate structures of the classroom that have supported this amplification in the next section as we trace the experiences of Nicole whose sense of relevance is also amplified through interactions between courses.

\section{"Nicole"}
Nicole is a genetics major who is interested in a career in data science in health informatics. Nicole has substantial disciplinary experience in the life sciences. She holds three jobs as an undergraduate: a biology laboratory learning assistant, a part-time researcher in an entomology lab, and a researcher in a genetics lab studying the effects of gene mutations on protein function. She has never taken physics before and has heard from her friends that physics courses at the university were difficult and terrible. Nicole decided to take our BLiSS Physics course after a recommendation from a friend who previously worked as learning assistant in the classroom. In addition to physics, Nicole is concurrently enrolled in biochemistry and a computational modeling course. We split our discussion of Nicole's experiences into two sections that focus on connections across settings in Nicole's mesosystem. The first section describes Nicole's views on the relevance of physics informed by connections between her coursework. We represent these forms of relevance in a way that aligns with previous cognitive perspectives used in physics education research \textemdash these views of relevance represent what the attitudinal and epistemological measures discussed earlier \cite{Redish1998a,elby1998epistemological,Halloun1998,Adams2006} aim to probe. The second section also looks at connections present in Nicole's mesosystem but pushes beyond the cognitive perspective into a situative perspective where we consider classroom structures, norms, and values as powerful agents that help co-construct and amplify relevance with Nicole.

\subsection{Nicole's perspective on the relevance of physics}
In this section we present evidence from a series of interviews to describe Nicole's beliefs around the relevance of physics. We focus on the mesosystem level to describe her perspective of how physics, computational modeling, and biochemistry courses can be relevant to one another.

\subsubsection{Relevance between computational modeling \& physics}
Our first glimpse that Nicole is seeing connections between physics and her computational modeling class is in our first interview in the early weeks of the Fall semester.\\

\begin{addmargin}[.45cm]{.45cm}
INTERVIEWER: Have you encountered physics in any of your other courses?\\
NICOLE: ...in my [computational modeling] class ---actually yesterday--- we had to come up with like, we ended up coming up with our own models for dropping a ball off of Beaumont Tower.  And then a skydiver so... and then I have homework to do for that class this weekend about a bungee jumper.  So yeah!  I am seeing physics a lot right now, actually.\\
I: So, what makes those examples physics to you?\\
N: For me, it's like whenever there's something involving motion, like I immediately think physics... So all those things that like with that [computational modeling] class, they're all involving some sort of motion, like actually they're all falling motions.  But, that is physics, like first thing that comes to mind. [Interview 1]\\
\end{addmargin}

Later in the same interview we probe if Nicole believes physics will be \textit{helpful} for completing her other courses or degree.\\

\begin{addmargin}[.45cm]{.45cm}
NICOLE: I've already seen it help me with like my [computational modeling] class just yesterday, actually.  Like, it was nice because I went from coding in physics' lecture.  And then I went to that, and we were doing like a similar type of thing.  So that was kind of nice to like it's nice to have those connected...\\
I: Wow. Can you tell me about some of these examples of ---so like take me into the perspective of you--- like when you walk into class, and after doing physics and now you're going to [the computational modeling course], when did you realize the basic connections?\\
N: Um, we were trying to like write some code, and I was like, "Well, hey?  Like I did this in physics today -- well, I'll be looking to see if there's any similarities, anything I could take from that."  And there was actually like one little point that I realized like, "Oh, I need to put this in my loop..." [Interview 1]\\
\end{addmargin}

Nicole is seeing parallels between the types of activities and resources used for both her physics course and her computational modeling course. The similarities between the type of activities cued her to \textit{bring in} elements from her completed physics solution into her computational modeling programming homework. \footnote{It should be noted again that these two courses were not designed intentionally to have this alignment and overlap; Nicole's experiences were the first indications that these two courses had the potential to be synergistic in promoting her sense of relevance.} Near the middle of the semester we follow-up with Nicole in a second interview and we ask her to elaborate on connections between physics and her other courses and again she makes connections to her computational modeling course.\\

\begin{addmargin}[.45cm]{.45cm}
NICOLE: Yeah, so I think it is important for me to have like a class like physics be able to connect to other classes because like I sometimes wonder like, "Oh, when will I ever use physics?" But like this [Wound Healing Activity \cite{Moore2014,Redish2014}] kind of shows that, "Oh, it is relevant". Like it does mean something to like what I might be wanting to do in the future. And that's another thing that like with another class I'm taking, I'm taking like that computational modeling. And sometimes I can kind of see how like physics would fit into that where at least like those concepts you use in physics that kind of seem to be applicable in that class, too. And we've done a lot of stuff with like modeling like the spread of disease and stuff. So that's been really interesting to kind of see how like even though it's not directly physics, it's like concepts that we've used in physics are also being used there.\\
INTERVIEWER: Gotcha. Do you have any specific examples of when you made those connections with the computational modeling class?\\
N: When we talked about in class yesterday, like the Lennard-Jones potential, like we ran a model like including like that specific equation.\\
I: Really?\\
N: Yeah. So it was modeling like- it was a lot like the one we did in [physics] class where we were modeling like the collisions. It was similar to that. And then like he used the Lennard-Jones potential... I was sitting in physics yesterday, and I like turned to my (inaudible), like, "Oh, hey, like we modeled this. I like have this equation on my computer right now that we used". So I thought that was cool.\\
\end{addmargin}

Nicole describes a moment where she is in her physics course and realizing that she has completed work from computational modeling that is relevant to the physics activity she's working on. This is analogous to the moment Nicole describes in the first interview, where she has a relevant programming activity from physics she can use to complete her computational modeling work. Physics and computational modeling are two settings that have a reciprocal connection that strengthens Nicole's belief that the concepts in physics are useful to computational modeling, and vice versa. Physics connecting to computational modeling has a deeper implication in that computational modeling is a critical skill for Nicole's planned future career in data science. In the above excerpt we see this connection between courses as entangled with Nicole's posited belief that physics may be relevant for what she wants to do in the future. In the next section, we continue to describe relevance within a cognitive perspective as we describe what can be considered an aim of true relevance, connections across two disparate disciplines: Physics and Biochemistry.

\subsubsection{Mesosystem Interactions Between Physics \& Biochemistry}

One of the learning objectives for the first semester of BLiSS Physics is to cover phenomena students may have seen in their Biology and Chemistry courses and to unpack the physics interactions underlying them. We see Nicole state that she may not use physics practices directly in a another course, but that physics explains why larger processes occur. We see Nicole apply this idea specifically to in-class activities in her Biochemistry course when we ask her to reflect on an activity exploring the force and energy required to unfold a protein using optical tweezers \cite{Redish2014}.\\

\begin{addmargin}[.45cm]{.45cm}
NICOLE: I thought this one was pretty interesting because I'm taking Biochem right now. So like we talk a lot about like protein unfolding and stuff and like the consequences it has. So like basically every lecture or so, she'll talk about like what happens with certain like mutations in genes and stuff. And a lot of it has to do with like messing with the protein structure. So I thought that was kind of interesting because it applied back to Biochem and kind of gave like an additional layer to what I had already learned in that class as to why stuff like that was happening. So that was interesting.\\
INTERVIEWER: So did knowing that like physics is involved in those processes help kind of...\\
N: It did actually kind of help like because like I understood the chemical background of it. But like seeing an actual like more of a physical reason for it rather than just, "Oh, like this amino acid is in the wrong place". Like it kind of like helped like connect; like bridge that gap between like, "Oh, this is in the wrong place. Because it's in the wrong place, this is happening". [Interview 2]\\
\end{addmargin}

Nicole states that physics provides a layer of understanding to the topics she covers in her Biochemistry class and finds these connections to help her bridge the gap between describing a process and understanding why it occurs. She revisits this idea later in the same interview when we ask her if she sees courses connecting to what she wants to do in the future.\\

\begin{addmargin}[.45cm]{.45cm}
NICOLE: Yeah, actually. All three of them [Biochemistry, Computational Modeling, \& Physics], I could see being applicable to like what I want to do in the future... with Biochem, like I can kind of see how like I want to kind of go in like a genetics-based type of informatics thing and like work on like bridging that gap between like genetics and the clinicians and how to help them. So like Biochem has helped with that because it's like shown me like different types of mutations and diseases and stuff and like why they're caused. And then the physics kind of goes with that because like Biochem shows why it's caused, and physics shows why the Biochem kind of happens. So yeah, I do think everything is kind of connecting.\\
INTERVIEWER: When did you first start thinking that the physics explains why the Biochem is happening?\\
N: It was these types of activities. I think there was another like similar protein type of one. And it was like that was when I kind of realized like, "Oh, like a month ago, we did this in Biochem, and now we're doing it in physics. It like makes sense". [Interview 2]\\
\end{addmargin}

In the past two sections we have discussed examples of Nicole making connections across courses she is enrolled in simultaneously. We now turn our attention to an instance where Nicole is able to continue making relevant connections after she has left the first semester of physics. 

\subsubsection{Mesosystem Interactions Between Physics \& Eukaryotic Cell Biology}

In our third interview with Nicole, halfway through the Spring semester, we discuss Nicole seeing physics come up in another course.\\

\begin{addmargin}[.45cm]{.45cm}
INTERVIEWER: Are you seeing physics pop up in other places?\\
NICOLE: I'm not seeing like physics as in like oh, this is physics. But more like concepts like in the Eukaryotic Cell Bio class I'm taking. We talked, we spent like, our first exam is on Tuesday, and so this first section was all about like transporting proteins and stuff. And so I know that a lot of it has to do with the structure of the structure of the transporters and stuff and we learned last semester [BLiSS] like that kind of hydrophobic interactions and stuff is important and it's because of physics, like physics.\\
\end{addmargin}

We see Nicole explaining that hydrophobic interactions are explained by the physics she's learned and that it has an impact on the structure and function of protein transporters. We again see Nicole pointing to physics as the mechanism behind why processes explored in her biology courses occur the way they do, but now she continues to make connections after she has left the first semester course. Nicole's statements describing the relevant connections between classroom activities in physics, computational modeling, biochemistry, and biology courses suggest that we can look to moments in her classroom experience in order to understand how relevant connections are built. In the next section we stay within the mesosystem to look for connections across the coursework Nicole experiences, but we push beyond the cognitive view that relevance is only constructed in Nicole's mind and consider the classroom's role in helping foster and amplify relevance.

\subsection{Situative perspective of the co-construction of relevance}
In the previous section, we described how Nicole's beliefs on the relevance of physics to her other courses has been impacted by her perception of classroom activities. We believe this view is in line with the kinds of relevance currently explored in PER and is reflected in the design of commonly used surveys. We argue, however, that Nicole's beliefs are not the only contribution to the relevance of physics. In order to capture a more full sense of this relevance, we turn our attention to the role the classroom environment plays in helping co-construct or amplify the relevance of physics to Nicole. In what follows, we employ a situative perspective to consider the reciprocal relationship of Nicole and her classroom in forging a sense of relevance.

\subsubsection{Relevance between computational modeling and physics}
13 days after we conducted Interview 2, Nicole and two other students are working in a group on an activity modeling the random walk of a protein colliding with water molecules by flipping coins. In the activity, the protein is said to move in one dimension either forward or backward depending on the result of the coin flip \cite{Redish2014}. When Nicole's group mates (Alisa \& Melanie, both pseudonyms) voice confusion and frustration, Nicole explains the point of the activity alluding to a piece of code produced in her computational modeling course that helped her understand.\\

\begin{addmargin}[.45cm]{.45cm}
ALISA: These probabilities are driving me nuts!\\
MELANIE: Is it the idea that it's wavering along this one line?\\
NICOLE: It's basically, you flip a coin... and if it's plus one if it's heads or minus one if it's tails... [my computational modeling code] makes it really easy to see.\\
\end{addmargin}

Shortly after this moment we see Nicole turn her laptop toward her group mates to show them the simulation she has written for her computational modeling course. The professor notices this moment while walking around the room and suspects that the group may be off-task. As the professor approaches, Nicole explains that she is showing the group a visualization she's written in her computational modeling course that relates to this activity. The professor asks to see what the visualization does, and we see Nicole show the professor and her group members her simulation run several times. Nicole explains how changing the number of flips the program simulates impacts the distance traveled in the random walk. The professor then smiles and asks if Nicole would be willing to share this with the entire class. In the last portion of the class, Nicole projects her laptop on the document camera for her classmates to see, and explains how her simulation creates a plot that helps visualize the randomness of random walk. She runs the code several times announcing how far the random walk ended relative to its initial position. Nicole explains that if you let the code simulate a large number of flips, there is a better chance of ending up further away from where it started. The professor then summarizes why she believes Nicole's code is a good visualization of the process of random walk.\\

\begin{addmargin}[.45cm]{.45cm}
INSTRUCTOR: So I thought this was really nice right? Because what Nicole is showing us is that, all she's showing you is that as the time increases, right? Which is the same as the number of flips, so number of flips is letting you go for a longer time, you are getting more likely to see a variation from zero. Right? So it's the same thing as that histogram plot. But this is showing it in a really nice way, like you can see where the zero is, right? And you can see where all of the variation that the steps gives you and where it ends up all in a single plot which was very nice.\\
\end{addmargin}

From the perspective of the ecological systems theory, this moment from the physics course embodies the reciprocal nature of participants and their settings. Nicole brings in and shares her computational modeling work as a way to show that it can make the purpose of this activity "really easy to see." The similarities between respective activities from computational modeling and physics mediate and support Nicole, but we also see Nicole impacting her course environment. The professor in this course invites Nicole to share her code with the class and summarizes why she believes Nicole's visualization is valuable in understanding the learning objectives for that activity. This incorporation of Nicole's contributions into the instruction suggests that her contributions have significantly impacted and augmented that day's physics lesson. In fact, after the class, author Nair interviewed the instructor who remarked that she initially had planned to switch to a planned demonstration, but revised her plans in-the-moment to make space for Nicole's contribution. The reciprocity in this moment \textemdash Nicole and her course environment impacting each other \textemdash is a critical part of the construct of relevance we want to describe and is in agreement with the features of Bronfenbrenner's (1979) ecological systems theory.

What sets this moment apart from the previous section is that Nicole's connection \textit{\textbf{is}} relevant to physics because she has become part of that days instruction. Her knowledge and previous work from a class outside of physics is brought in, validated, and incorporated into what it means to do physics. If the classroom design did not include the flexible time and space to allow for Nicole to bring her expertise \textit{\textbf{in}}, then we may not have seen this moment play out as we had. 

Stating Nicole \textit{\textbf{believes}} physics is relevant due to these interactions is not capturing the whole of the moment. The instructor, through her actions, suggests that Nicole's contributions are relevant and useful. The classmates that listen and give Nicole their attention signal what is happening is relevant to their learning of physics. The technology utilized in the classroom allows Nicole to project her personal computer makes this type of sharing-out possible; this type of student-led presentation is a critical piece of the course's design. It is all of the classroom roles, structures, norms, and values in concert that allow for this sense of relevance to be co-constructed and amplified. 

In order to connect this expanded view of relevance back to the belief structures mentioned before, we return our focus back onto Nicole's beliefs to verify that this moment was \textit{\textbf{relevant}} to her. Author Nair, after seeing this moment play out in the classroom, followed up with Nicole over email. The following exchange happens in the very same day the events took place.\\

\begin{Email}{gray}{\Abhilash}{\Nicole}
Hi Nicole,\\

I had walked into class and noticed you were presenting work from a Jupyter notebook. I'm wondering if you could reply with a brief reflection of you presenting that to the class:\\
 - how did that happen?\\ 
 - how did it feel?\\
 - what are your thoughts about presenting other work in your physics class?\\

I just want to grab your thoughts while the memory is still fresh in your mind.\\ 

Have a wonderful day!
\end{Email}
\begin{Email}{gray}{\Nicole}{\Abhilash}
Hi Abhilash!\\

When we started the coin flip/random walk worksheet, I realized that I had essentially done this in my [computational modeling] class, so I showed my group, because I thought it was a lot more easy to visualize like that than it was to physically flip a coin. [The professor] saw me doing it and asked if I would show the class. I actually enjoyed doing it, because there have been a lot of connections between the two classes, so it was fun to be able to show others the connections I had been able to make. I think showing work from other classes actually helps a lot, because it puts into perspective all the different ways physics is used.\\

Hope you have a great day!
\end{Email}

Nicole believes the act of bringing in work across courses is helpful and she enjoyed showing her classmates the connections she was able to make. In the next section we look at a connection between biochemistry and physics Nicole recalls in her third interview and highlight the critical features that helped support this memorable and relevant moment.

\subsubsection{Relevance between biochemistry and physics}
In our third interview with Nicole, we find out that connecting across physics and biochemistry is a lasting memory for her.\\

\begin{addmargin}[.45cm]{.45cm}
INTERVIEWER: Tell me some memorable moments from the first semester.\\
NICOLE: Um I think one memorable moment was like towards the end when we were talking about like the water molecules and stuff, I really liked that because I felt like it related really well to Biochem, so I thought that like those two concepts like together like really gave me a really strong idea of why polar and non-polar molecules don't interact basically.\\
I: Oh, ok, so take me to the phys-- which parts were they when you talked about water.\\
NICOLE: Um that was the very end of the semester we basically just did-we didn't really do any like experimentation or anything we just kind of talked through a lot of it and made like giant whiteboards and drew a lot of pictures and stuff. Um but I remember like there was one specific question where I was like "we literally had an entire slide in my biochem lecture that is this word for word" So I like showed them and I'm like this is what's going on and this is like how the physics would work behind it so.\\
I: Whoa, so you were showing your group-mates?\\
N: Yeah\\
I: In physics?\\
N: Hm.mmm. [nods in agreement]\\
I: Um it seems like a lot of times you're bringing in outside of class materials into the class, how did that feel?\\
N: It was nice, it was nice to make that connection between the two, because it kind of like reinforces the idea that science is super interconnected.\\
\end{addmargin}

The activity which Nicole refers to above is adapted from NEXUS/Physics which asks students to use concepts of enthalpy and entropy to reason through why oil and water do not mix. It explicitly questions why oil and water mixed together isn't a favored state even though it appears more "disordered" \cite{Redish2014}. We now turn to the in-class video to see this moment play out. Nicole is working in a group of three with classmates Melanie and Alisa; Nicole and Melanie are adjacent to each other. All three students have their heads down filling out answers to first page of the activity. Nicole finishes answering the first page and turns the page to look at the next question (\#4). She then exclaims "Ooh! Number 4 I..." and smiles. She then quickly reaches for her laptop and powers it on. A short while later, Melanie is finished with the first page as well and confirms her answers with Nicole. She then notices Nicole's laptop screen and asks Nicole "It's biochem?" Nicole then replies with a smile and says "Yup, but it's literally the answer to number 4, so..." Nicole notes multiple times to classmates and to undergraduate learning assistants that the concepts in this activity are being covered simultaneously in her biochemistry course. 

One learning assistant (Miles, pseudonym), who is coincidentally majoring in biochemistry, stops by Nicole's group. Miles is a case student from Year 1 of BLiSS Physics, who joined the instructional team as a learning assistant in the following year. Miles validates Nicole's connection and states that a similar realization happened for him when he was a student in the course; he had seen similar diagrams being used across physics activities and his biochemistry textbook. He, again relays his own experience by describing his former belief that the processes described in the activity were simply driven by enthalpy and describes how he now realizes he was wrong. When Nicole and her group members suggest that they are finished, Miles challenges them with a task that is not part of the activity: to consider if the process of a substrate binding to an enzyme is favorable. Through this interaction, Nicole making connections across her courses is normalized and reaffirmed by a learning assistant who has the dual status of (1) a student who has successfully completed the physics course and (2) has disciplinary expertise in biochemistry.

Over the course of two class periods we see Nicole taking the lead on writing her group's answers on white-boards, this was not common for Nicole across her entire semester in the course. Working with Melanie and Alisa was a stated preference of Nicole's and she reported that she worked much more effectively with them than other students. Whether with this group or others, Nicole typically did not take the lead on writing her group's responses and instead can be often seen on video quietly following her group. Similar to the synergy between Nicole's computational modeling course and physics, the overlap between biochemistry and physics activities was not planned. We argue that the similarity of activities is not the critical feature in making this moment memorable and relevant. 

The data from the in-class video suggests that there are several classroom structures that help facilitate Nicole bringing in relevant biochemistry knowledge into a physics activity. The classroom norms and culture encourage students to seek out any and all resources they require. Many students use their own laptops, phones, and tablets to look up information or read class materials online; this is encouraged and not considered off-task. Nicole projects her laptop to the class to show her computational simulation of random walk and uses her laptop to bring up her biochemistry notes. The classroom culture prioritizes student ideas and building consensus over relaying the "correct answer." This affords Nicole the opportunity to take the lead and demonstrate her knowledge from biochemistry and contribute to her group's white-boards. Miles's validation of Nicole's connections are bolstered by his roles as a senior biochemistry major and a student who has successfully completed the physics course. 

All of the classroom structures, norms, and roles that help facilitate and co-construct this moment cannot be described by simply measuring Nicole's beliefs. We argue that the full sense of physics being relevant to Nicole's other courses exists in her beliefs as well as the interactions between Nicole and the classroom. Pre-to-post shifts on attitudinal and epistemological surveys can be an important data point for measuring relevance, but it does not capture the richness and the power of classroom elements in amplifying or co-constructing those beliefs. Using ecological systems theory allows us to identify that the mesosystem can serve as a rich space for relevance through connections across courses. Our situative approach to relevance gives us the chance to (1) capture these moments as they play out in our classrooms for a more complete sense of relevance and (2) highlight course structures and norms that be designed/iterated to impact the relevance of physics for students.

\section{Discussion}
In this paper, we have presented analysis that pushes beyond the cognitive perspective on relevance and uses a situative approach \cite{Greeno1996}, a reader may question if exploring relevance necessitates such an expansive framework. Ecological systems theory's situative nature gives scholars the language to start identifying sites that may be formative in a student's sense of the relevance of physics. Until now, the physics classroom has been the primary focus of work on students' beliefs on the relevance of physics \cite{Adams2006,Redish1998a,Halloun1998,elby1998epistemological,Stuckey2013}. We argue that the other layers of systems within a student's life can play a critical role in a student's sense of relevance. In addition to allowing us to map out important contexts in students' lives, a situative perspective affords us the ability to explore the reciprocal relationship that an environment and its participants have on one another.

\subsection{Relevance is co-constructed by students and their environments}
One of the limitations in seeing relevance as purely a cognitive construct is that it puts the onus on students to \textbf{\textit{believe}} that physics is relevant. When students report their belief that physics may not be relevant to their lives, there is a risk that scholars may focus on how to fix students' beliefs as opposed to addressing the layers of systems students exist within. Moving away from a deficit-view of students beliefs, we can explore the rich ways students and physics classrooms can co-construct and amplify the relevance of physics.

Maria arguably is a student that will attempt to make connections in any classroom she is in, but it is the dynamic relationship of the two courses that helped forge her sense that physics is relevant to her. The curriculum \cite{Brewe2008,Redish2014} may have introduced the notion that physics and biology share disciplinary ideas, but we argue that relevance was co-constructed by Maria, her coursework, the roles she played in those settings, and the roles the settings played in amplifying her connections. This reciprocal and dynamic nature to the construction of relevance is well suited for a situative perspective. Maria is able to bring her biology knowledge into the physics classroom to arrive at an alternative solution path. The instructor validates Maria and asks her to share her reasoning to the rest of the class. The course allows for multiple correct answers and has the space for students' voices to be shared as experts in their home disciplines. In this moment, Maria is bringing in the outside disciplinary knowledge, but it is the classroom structures and values that amplify her relevant connection. 

Nicole sees connections between her physics course and her computational modeling course early on in the semester. This connection is supported by multiple moments in which Nicole is able to bring in the work from one course to help with the other. We demonstrate through a series of interviews that Nicole's beliefs around of relevance have been impacted by her classroom experiences. When we look to the classroom for evidence of this change, ecological systems theory allows us to see the role the classroom plays in co-constructing and amplifying the relevance of physics for Nicole. Nicole and the classroom's design together have constructed a moment in which physics and Nicole's disciplinary knowledge and experiences are relevant to each other.

\subsection{Ecological systems theory as a motivation for holistic reforms}
One of the strategies employed by scholars attempting to improve the results from survey measures is to engage students in activities to develop their abilities in reflection and meta-cognition \cite{Elby2001a,Bennett2016}. These interventions are often localized to the physics classroom and actively require students to articulate and reflect on connections between physics and their lives. The tacit message is that students do not already possess adequate abilities in meta-cognition and reflective practices. This deficit-view of students' abilities is problematic, but it becomes especially so when considering non-physics majors. The life science students presented here are high-achieving and have sophisticated expertise in their major disciplines. They actively and regularly engage in meta-cognition to assess how relevant their coursework and other experiences are to them, including deciding whether physics is a course that should be taken simply "to get it out of the way."

It may be possible that a physics classroom activity requiring students to construct and reflect on potential connections of physics to their life can result in further development of students' sense of relevance of physics, which can then be captured in pre-to-post shifts on surveys. It may also be possible that repeated insistence that students should believe that physics is connected to their lives results in them reporting what they have been conditioned to report. Without careful qualitative work, it will be difficult to distinguish between these possibilities. We acknowledge that a positive result on these survey measures regardless of the strategy is probably beneficial for the perception of physics education. The aspect that concerns us is that as physics reforms proliferate across institutions, we continue to collect instructional strategies and activities to layer on top of a physics classroom and lose sight of the ecology at play. Students enter and exit a variety of disciplinary classrooms and without a systems-level view of their ecosystem, it is challenging to forge meaningful connections that persist and perpetuate beyond the physics classroom.

Ecological systems theory motivates us to move beyond attempts to design or choose the best activity to promote students' beliefs around physics. Instead it necessitates that we see students as existing in multiple interacting layers of systems. In our course, we have established several practices to ensure that students beliefs, affect, and experiences are brought into the instructional cycle. We use an initial survey to assess which disciplinary areas our students are coming from, what they hope to do in the future, their previous experiences with physics, and to gauge their level of anxiety or fear as they enter the course. This serves as a foundation of knowledge that will help guide instructional choices, but it must be noted that even with this knowledge we did not predict the connections Maria and Nicole were able to make. Planning instruction to attend to students' ecologies is less about predicting which activities will resonate and more about designing activities with space, depth, and flexibility to amplify what students are already capable and willing to do. This process promotes bringing student ideas into activities, allows instructors to position life science students as experts in their home disciplines, and provides critical validation and affective support. Moving toward holistic physics reforms that are responsive to the ecosystem present for their students has the possibility to generate long-lasting and meaningful narratives of the relevance of physics to students' lives.

\subsection{The importance of providing space for students to bring the whole of their disciplinary selves in}
One of the components of the students experiences outlined in this paper is the space they are afforded in bringing in their disciplinary expertise. We argue that the participatory acts these three students engaged in was critical in the construction of relevance. Maria recognized an alternative solution path utilizing her microbiology content knowledge and was afforded the opportunity to share her solution to the class. Nicole brought in her solution from her computational modeling course to show her group mates a visualization that she felt was easier to understand than the physics activity, she was invited to share this with the entire classroom. These participatory structures and a flexible content coverage schedule were important considerations to the design of our classroom. Without this responsive and dynamic environment, we imagine Maria and Nicole m may not have had as positive of an experience.

\section{Implications}

\subsection{Revisiting our tools to measure relevance}
A large portion of the work on students' attitudes and beliefs in physics education research consists of reporting on four commonly used surveys. This corpus of work focuses on evaluating curriculum and pedagogy on its ability to shift students' endorsements of belief statements. This work has expanded to include differing student compositions and has been used to compare the relative success of different research backed instructional strategies. Results from these surveys have have been frequently reported to suggest that it is common and expected that students will find physics less relevant after instruction than they did before. We argue that the first issue in this chain of events is that there is no consensus for what it means for physics to be relevant. In the surveys, phrases such as "real world" and "everyday life" are used without contextualizing them into what students actually experience. We have previously argued that the concept of the "real world" presented in these surveys is of an abstracted sense of the natural world that does not include students' disciplinary interests \cite{Nair2018}. If our physics students experiences are not appropriately captured within these surveys, then the measurement of relevance is limited and the interpretations of students' abilities to make connections becomes problematic. We still see the benefit of pre-to-post surveys and will continue to employ them to understand the effects of our classrooms, but we must acknowledge the limitation in their ability to capture relevance. Most importantly, we must shift away from describing students in deficit-tinged language so that the role of instruction is not to persuade, demand, or force students to make connections but rather invite students to have a role in the classroom so that physics may have a role in their lives.

\subsection{Expanding the goals of designing courses for relevance}
The reader may wonder if the solution to creating more relevant physics classrooms is to design the next coin flip activity for the next Nicole, or the next paramecium activity for the next Maria. Much of the curriculum development in introductory physics for the life sciences (IPLS) is focused on creating interesting and engaging activities for students to authentically apply physics to explore biological phenomena. We believe that is certainly an important factor in promoting relevance; activities must be rich and authentic as to not simply provide a veneer of biological context over what is essentially a traditional physics problem \cite{Gouvea2013c}. We also believe that the curriculum is only part of the solution; classroom structures, norms, and roles play an equally, if not, more important role. 

The data presented in this paper from the experiences of Maria and Nicole show the importance of the classroom in bringing in and valuing their disciplinary experiences. We are finding that providing space for students to bring their ideas into the classroom requires a relatively flexible content coverage schedule, norms for sharing out ideas to the larger class, and facilitation of the building of consensus. These are elements of the classroom that can be intentionally designed. As part of our larger design based research study, we are currently in the process of articulating design conjectures for a classroom that among other things, can amplify relevance. We do not believe it is possible to predict all of our students' disciplinary interests and experiences, so we rely on building space to allow for students to share their expertise. This is important for both for students and learning assistants who were former students. Positioning students as experts in the life sciences and incorporating their expertise into the learning of physics allows for a responsive course and is less reliant on a perfectly coordinated IPLS curriculum.

\section*{Acknowledgements}
We thank the members of the ANSER group for thoughtful discussions around this work. We thank the PERL@MSU group for their engagement and reflections during group meetings where this work was presented. This work was funded by the Lyman Briggs College and the Department of Physics and Astronomy at Michigan State University. We would also like to thank Angela Little and Benjamin Geller for their encouragement and constructive feedback in pushing this work forward. Lastly, and most importantly, we want to thank Beverly, Miles, Maria, and Nicole for participating in our research and allowing us to share their experiences.

\bibliography{Mendeley.bib}
\end{document}